\documentclass[draft, sw]{agutex2015arx}

\usepackage{lineno}
\usepackage{graphicx}
\usepackage{amsmath}
\usepackage{amsfonts}
\usepackage{amssymb}
\usepackage{amsthm}
\usepackage{url}

\newtheorem*{theorem}{Theorem}

\setkeys{Gin}{draft=false}
\authorrunninghead{TSIFTSI, T. and DE LA LUZ, V.}

\titlerunninghead{EXTREME SOLAR FLARES}

\authoraddr{Corresponding author: Thomai Tsiftsi,
Centro de Ciencias Matem\'{a}ticas, Unidad Michoac\'{a}n, Universidad Nacional Aut\'{o}noma de M\'{e}xico, Morelia, Michoac\'{a}n, CP 58190, M\'{e}xico
(ttsiftsi@matmor.unam.mx)}

\begin{document}
\onecolumn
\title{Extreme Value Analysis of Solar Flare Events}

\authors{T. Tsiftsi\altaffilmark{1} and V. De la Luz\altaffilmark{2}}
\altaffiltext{1}{Centro de Ciencias Matem\'{a}ticas, Unidad Michoac\'{a}n, Universidad Nacional Aut\'{o}noma de M\'{e}xico, Morelia, Michoac\'{a}n, CP 58190, M\'{e}xico}
\altaffiltext{2}{Conacyt, Laboratorio Nacional de Clima Espacial, Instituto de Geof\'{i}sica, Unidad Michoac\'{a}n, Universidad Nacional Aut\'{o}noma de M\'{e}xico, Morelia, Michoac\'{a}n, CP 58190, M\'{e}xico}

\keypoints{\item Analysis of peak flux from solar flares
\item Application of extreme value theory
\item Modelling by generalised Pareto distribution.
}


%
%


\begin{abstract}
  Space weather events such as solar flares can be harmful for life and infrastructure on earth or in near-earth orbit.
In this paper we employ extreme value theory (EVT) to model extreme solar flare events; EVT offers the appropriate tools for the study and estimation of probabilities for extrapolation to ranges outside of those that have already been observed. In the past such phenomena have been modelled as following a power law which may gives poor estimates of such events due to overestimation.  The data used in the study were X-ray fluxes from NOAA/GOES and the expected return levels for Carrington or Halloween like events were calculated with the outcome that the existing data predict similar events happening in 110 and 38 years respectively.   
\end{abstract}

\begin{article}

\section{Introduction}
Space weather effects such as solar flares, geomagnetic storms and solar energetic particles can be hazardous to human activities, both on earth and in near-earth orbit, and the scientific community is faced with a number of such phenomena of great variability \citep{Riley2017}. A probabilistic assessment of the likelihood of such events -- and their strength -- is a top priority in space weather forecasting \citep{2017SpWea..15..282J}. In particular, the prediction of occurrence of solar flares is crucial as even small intensity events can adversely affect geopositional systems and telecommunications \citep{Lanzerotti2007}. The immediate effects of solar flares are mainly related to technology i.e.  telecommunications, radio transmissions, and satellites. X-ray emissions can even affect and disturb the operation of radars and other devices. For example, during the Halloween storm of 2003 a set of perturbations related to the arrival of Coronal Mass Ejections at Earth were recorded: satellite communications, auroras from Texas to the Mediterranean, and several disruptions in the transport of high voltage energy (GICs) that produced power cuts \citep{watson121intense}. At long term and middle latitutes, high-voltage transmission networks are also affected by Space Weather events \citep{2012SGeo...33..973D,SWE:SWE6}.

\bigskip
\noindent
Solar flares are classified with letters: A, B, C, M, and X according to the peak flux measured in Wm$^{-2}$ of X-rays as measured by the Geostationary Operational Environmental Satellite (GOES) spacecrafts. Each class of events has subclasses ranging from 1 to 9. Each letter  specifies a flare of one order of magnitude larger than the previous one. For example an M2 is $2 \times 10^{-5}$ Wm$^{-2}$ whereas an X2 is $2 \times 10^{-4}$ Wm$^{-2}$. Furthermore, an X2 is twice as strong as an X1 whilst four times more powerful than an M5 flare \citep{XRS}. 

\bigskip
\noindent
To forecast events that are of interest to human activity it is useful to study the distribution of the strengths of events recorded by the satellites mentioned above. Of particular interest is the study of the tail of the distribution of solar flares which describes the occurrence of extreme events such as the Carrington event of 1859 \citep{Carrington}. Up to today it has been assumed that the tail of the distribution of flare strengths $x$ follows a power law \citep{Riley, Lu} i.e. $\mathbb{P}(X > x) \propto x^{-a}$, with values of $a$ ranging between 1.7 and 2 \citep{Boffetta, Aschwanden} or a lognormal distribution \citep{Riley2017}; however it is believed that a power law distribution overpredicts extreme events and their occurrence. In addition, in \citep{Riley} as 
mathematical and numerical discrepancies have been noted \citep{Parrott1, Parrott2, Roodman}, the confidence intervals of the predictions are not reported \citep{Riley2017} and fits of straight lines on curved distributions overestimate the predicted risks as straight lines can be good approximations locally but cannot be accurately used for extrapolating to extremes.

\bigskip
\noindent
In this paper, we suggest more appropriate tools for the study of extreme space weather events for X-ray flares of Carrington's magnitude or more. We suggest the study of extreme solar flare events with the employment of Extreme Value Theory (EVT) for the estimation of the probability of extreme solar flares. The main result of this paper is that a Carrington-like event (X45 $\pm 5$) \citep{Cliver} is expected approximately once every 110 years whereas a Halloween-like event (X35 $\pm$ 5) \citep{Cliver} is expected approximately once every 38 years. Furthermore, the frequencies of these events happening in the next solar cycle are in very close agreement with the NOAA predictions.

\bigskip
\noindent
The paper is structured as follows: in Section \ref{EVT} we introduce the methodology and the motivation behind EVT as well as its different approaches; in Section \ref{data}  we describe the data set used throughout this paper and discuss some modelling issues relevant to the dataset. In Section \ref{results}, we implement EVT and present the results and in Section \ref{seasonality} we evaluate the effect of the seasonality inherited from the sun in our predictions. In Section \ref{conclusions} we conclude with a discussion about our findings.

\section{Extreme value theory}
\label{EVT}
Extreme value theory (EVT) is a branch of statistics that offers tools and techniques for the study and estimation of probabilities of events outside the range of those that have already been observed. Since these extreme events are sparse their estimation requires extrapolation beyond the observed levels. EVT is designed precisely for such extrapolation and utilises asymptotic analyses for the foundation of extreme value models. Hence, EVT can model the stochastic nature of process of unusually large or small intensity events \citep{Coles}. What is remarkable about EVT is that it avoids any assumptions about the underlying probability distributions: the possible asymptotic distributions for extreme events are universal, independent of parent (or true distributions) describing the full process and thus mitigate the necessity of making any a priori assumptions. Hence, EVT is the most rigorous general approach allowing us to extrapolate from the past.

\bigskip
\noindent
EVT has been previously used to model a range of natural phenomena such as extreme wind speeds \citep{Fawcett}, extreme draughts and flooding \citep{Canfield, Katz} and is has even been used for public health and the prediction of extreme outbreaks of pneumonia and influenza \citep{Thomas}. In the area of space weather, EVT has been used to estimate the disturbance storm time (Dst) index \citep{Tsubouchi, Silrergleit}, the daily $\text{A}_a$ index \citep{Silrergleit2}, speeds of fast Coronal Mass Ejections (CMEs) \citep{Ruzmaikin}, the magnetic index $\text{A}_p$ \citep{Koons}, energetic electron fluxes \citep{Koons, Obrien}, relativistic electron fluxes \citep{Meredith} as well as the millennial sunspot number series \citep{Acero}.

\bigskip
\noindent
EVT is based on two main approaches: 1) the block maxima approach (also known as the Fisher-Tippett-Gnedenko theorem) and 2) the peaks-over-threshold approach (also known as the Pickands-Balkema-de Haan theorem \citep{Balkema, Pickands}) \citep{Coles}. In this paper we suggest the use of the latter in our analysis for reasons explained in Section 4. 

\subsection{Block maxima}
The block maxima approach consists of grouping the observed data into $N$ blocks of equal length $n$ and extracting the maximum of each block. Then, according to EVT \citep{Coles} the distribution of maxima in the large $n$ limit is a renormalised Generalised Extreme Value (GEV) distribution. Let $X_{1}, X_{2}...,X_{n}$ be a sequence of independent and identically distributed (i.i.d) random variables with common distribution $F$. Then, it can be proven that the cumulative distribution function (CDF) of $M_{n}=\text{max}\{ X_{1},...,X_{n}\}$ is:

\begin{align}
\mathbb{P}\left( \frac{M_{n}-b_{n}}{a_{n}} \leqslant z\right) \longrightarrow G(z), ~~~~\textrm{as n} \longrightarrow \infty
\end{align}

\noindent
if constants $\{a_n >0\}$ and $\{b_n\}$ can be found for the limit to exist\footnote{The constants $a_n$ and $b_n$ are chosen in order to allow a linear renormalisation of the variable $M_n$ so that the distribution of $M_n$ does not degenerate to a single point in the large $n$ limit. Appropriate choices of $a_n, b_n$ standardise and stabilise the location and scale of $M_n^* = \frac{M_n-b_n}{a_n}$ as $n$ increases. Explicit expressions for both $a_n$ and $b_n$ can be found in \citep{Castillo}, page 204.}, where $G$ is a non-degenerate distribution function which is a member of the GEV family:

\begin{align}
G(z) = \exp \left\{ - \left[  1+ \xi \left( \frac{z-\mu}{\sigma}\right) \right]^{-1/ \xi} \right\}
\end{align}

\noindent
defined on \{$z: 1 + \xi(z-\mu)/ \sigma >0$\} and where $-\infty < \mu < \infty $, $\sigma>0$ and $-\infty <\xi<\infty$. 

\bigskip
\noindent
The GEV family of distributions has three parameters, the location parameter $\mu$, the scale parameter $\sigma$ and the shape parameter $\xi$. Estimators for the three parameters can be found by the Maximum Likelihood Estimate (MLE) method. Maximum  likelihood estimators maximise the probability of observing the recorded data given a parameter dependent model. MLEs for the three parameters must be found numerically as there is no closed form solution. The GEV can be categorised according to the value of the parameter $\xi$ as follows: if $\xi>0$ then the distribution is Frech\'{e}t (``heavy'' tail), if $\xi<0$ the distribution is Weibull with an upper bound at $\mu-\frac{\sigma}{\xi}$ (bounded tail) and in the limiting case of $\xi \rightarrow 0$ the distribution is Gumbel (``light'' tail).

\bigskip
\noindent
However, in this approach there are many drawbacks that cannot be easily overcome. For example, the choice of the block length is crucial. Although it is often convenient is to choose a block length of one year i.e. yearly maxima, this is not always the optimum choice for all datasets, and is not the most natural length for solar flares. The choice is usually the block length that balances the variance and the bias. Furthermore, since extreme events are scarce this method is not ideal as it wastes a lot of good data when the block size is large. This motivates the peaks-over-threshold approach. 

\subsection{Peaks-over-threshold (POT)} 

A different yet related way to determine which observations are extreme is the peaks-over-threshold (POT) approach where \textit{all} observations greater than some high threshold are deemed extreme. The POT approach is used to model the tail of the distribution above the chosen threshold and the theory states that for a sufficiently high threshold $u$ the conditional distribution of exceedances is a Generalised Pareto Distribution (GPD) \citep{Coles, Leadbetter}.

\bigskip
\noindent
To make this more concrete, let $X_{1}, X_{2}...X_{n}$ be a sequence of i.i.d random variable with $M_{n}=\text{max}\{ X_{1},...,X_{n}\}$. If in the large $n$ limit, $\mathbb{P}\left( \frac{M_{n}-b_{n}}{a_{n}} \leqslant z\right) \approx G(z)$ with $G(z)$ following a GEV then for large enough $u$ the distribution of $Y= X-u$ conditional on $X>u$, is approximately

\begin{linenomath*}
\begin{equation}
H(y) = \mathbb{P}(Y \leqslant y | Y>0) = 1- \left(1 + \frac{\xi y}{\tilde{\sigma}} \right)^{-1/\xi} 
\label{pareto}
\end{equation}
\end{linenomath*}

\noindent
defined on $\{y: y>0$ and $(1+ \xi y / \tilde{\sigma}) >0\}$ where $\tilde{\sigma} = \sigma + \xi (u - \mu)$. The parameters of the distribution are similar to the GEV with $\xi, \mu, \sigma, u$ the shape, location, scale and threshold respectively. Equation \ref{pareto} is the Generalised Pareto family and it implies that if block maxima are approximately GEV distributed then the threshold excesses are approximately GP distributed.

\bigskip
\noindent
The issue of choosing a threshold is similar to the objectivity of the choice of block size in the GEV approach. Similarly, the choice should be one that assures balance between bias and variance. There are two ways to assess the suitability of the threshold: the \textbf{mean residual life plot}  as well as the  the stability of the parameters' MLE estimates across a variety of different thresholds. For the former, if the threshold excesses follow a Pareto distribution then $\mathbb{E}(X-u|X>u)$, the mean of the excesses of the threshold $u$, should be a linear function of $u$ at levels for which a Pareto model is appropriate. A good and unbiased estimator of the mean is the sample mean thus one should expect it to change linearly with $u$ (within confidence intervals). For the latter, a complementary technique is to fit the GPD at a range of $u$ and look for stability of the GPD's parameters, again within confidence intervals. If a GPD is appropriate for excesses of a threshold $u$ then excesses of a higher threshold $u'$ should also follow a GPD with exactly the same shape and scale parameters. However, sampling variability does not guarantee exact stability but the parameters should be stable within their sampling errors. Plotting both $\xi$ and $\sigma$ against $u$ with their confidence intervals, $u$ should be selected as the lowest value for which the estimates remain near-constant.

\bigskip
\noindent
As with the GEV, and due to the duality between them, the shape parameter of the GPD determines its tail's qualitative behaviour. Thus, $\xi>0$ yields the Pareto CDF (``heavy'' tail), $\xi<0$ yields the Beta CDF (bounded tail) with upper bound at $u-\frac{\sigma}{\xi}$ and the limiting case $\xi \rightarrow 0$ yields the exponential CDF (``light'' tail).

\bigskip
\noindent
Although there are several choices to estimate the parameters of the fitted GPD, in this paper we use the Maximum Likelihood Estimation (MLE). The estimated values of $\xi$ and $\tilde{\sigma}$, can be found by maximising the likelihood function; in fact, it is standard practice to maximise instead the log-likelihood. Assuming that ${y_{1},...,y_{n}}$ are the $n$ exceedances above $u$, with $y_{i}=x_{i}-u$ conditional on $X>u$ then the log-likelihood function to be maximised provided that $\xi \neq 0$ is:

\begin{linenomath*}
\begin{equation}
\mathcal{L}(\xi,\sigma) = -n\log(\sigma) - \left( 1+\frac{1}{\xi}\right) \sum_{i=1}^{n} \log\left( 1+\frac{\xi y_{i}}{\sigma}\right)
\end{equation}
\end{linenomath*}

\noindent
provided that $(1 + \sigma^{-1}\xi y_{i})>0$ for $i = 1,...,n$ otherwise $\mathcal{L}(\xi,\sigma) = - \infty$. However, as in the GEV case, there is no analytic solution for the maximisation of the likelihood so numerical iterative, optimisation methods are used. Along the estimates, we can obtain their standard errors and confidence intervals by using standard techniques (see \citep{Coles, Castillo}).

\subsection{Discussion between approaches}

What is really remarkable about both aforementioned methods, is the fact that the three GEV distributions, and hence the three types of the GPD family, are the only possible limits for the distribution of maxima $M_n$ \citep{Coles}, no matter what the parent distribution $F$ of the population is. This result can be seen as an extreme value theory analogue of the central limit theorem. This in turn shows that both the GEV and GP distributions are the appropriate limits for the two approaches when the number of samples becomes increasingly large \citep{Reiss}.

\bigskip
\noindent
The two mentioned methodologies are strongly connected: if block maxima obey the GEV distribution then exceedances over some high threshold obey an associated GPD. Furthermore, the shape parameters of the two distributions are expected to be identical asymptotically \citep{Leadbetter, Coles}. Although this argument stands in theory and the two approaches are equivalent in the limit of infinitely long time series, the GP distribution is more robust and provides more realistic results when finite time series are considered \citep{Ding}. Furthermore, as an approach it does not waste so many valuable extreme data as the block maxima, as it is known that extreme data are scarce and very hard to observe or collect. Thus, the GPD approach is the one that we will employ in this paper.

\section{Data}
\label{data}

The data used in this study
were extracted from the SWPC/NOAA website and represent solar flares normalised from H$_{\alpha}$ spectral observations and X-ray fluxes spanning a period of 43 years from November 1975 to July 2017 (\url{https://www.ngdc.noaa.gov/stp/space-weather/solar-data/solar-features/solar-flares/x-rays/goes/xrs/}). In order to take into account the last energetic events of the current solar cycle, we also included the records from July 2017 to October 2017.

\bigskip
\noindent
The analysis was run on the peak X-ray flux of each solar flare event converted in $Wm^{-2}$. The data have been scaled in order to yield consistency and long term continuity across all different GOES satellites that have been recording data since 1975. To get the true X-ray fluxes for the GOES satellites the data were divided by 0.7 which ensures that the classification level of the solar flares is consistent across all GOES \citep{Machol}.

\bigskip
\noindent
The POT method is valid for independent and identically distributed (iid) data which we verify with post analysis diagnostics. Since the analysis is done on the peak X-ray fluxes only, the data are assumed to be independent; any background noise or any consecutive events that are usually recorded multiple times in the GOES data, as most events last for a long time, have been removed by choosing the peak of each event only. This statement is further supported by the estimated extremal index of the data: for this we state the following theorem. 

\bigskip
\bigskip
\bigskip
\noindent
\textbf{Extremal index}
\begin{theorem}

\bigskip
\noindent 
Let $\tilde{X_{1}},...,\tilde{X_{n}}$ be a stationary series that satisfies Leadbetter's condition \citep{Coles, Leadbetter}, i.e. showcasing sufficiently weak long-range independence, 
and let $\tilde{M_{n}}=\textrm{max}\{\tilde{X_{1}},...,\tilde{X_{n}}\}$. Now let $X_{1},...,X_{n}$ be an independent series with $X$ having the same distribution as $\tilde{X}$ and let $M_{n}=\textrm{max}\{X_{1},...,X_{n}\}$. Then if $M_{n}$ has a non-degenerate limit law given by $\mathbb{P}(M_{n}-b_{n}/a_{n}  \leqslant x) \rightarrow G(x)$ it follows that:

\begin{align}
{P}\left\{ ( \tilde{M_{n}} - b_n/ a_n  )\right\} \rightarrow G^{\theta}(x) 
\end{align}

\noindent
for some $0 \leqslant \theta \leqslant 1$. The parameter $\theta$ is known as the extremal index and quantifies the extremal dependence with $\theta =1$ a completely independent process and $\theta \rightarrow 0$ demonstrates increasing levels of dependence. 

\end{theorem}

\bigskip
\noindent
For the dataset used in this study and to reinforce the statement of independence above, the estimated extremal index of the data used, as estimated by the runs declustering method \citep{Coles, extremes}
, was found to be $\theta = 1$. We conclude that our data are completely independent.

\bigskip
\noindent
Seasonality is an important aspect to consider when analysing time series. We analyse possible seasonality that could be inherited from the solar cycle in Section 6.

\section{Results}
\label{results}

\begin{figure}[h!]
\centering
\includegraphics[width= 0.5\textwidth]{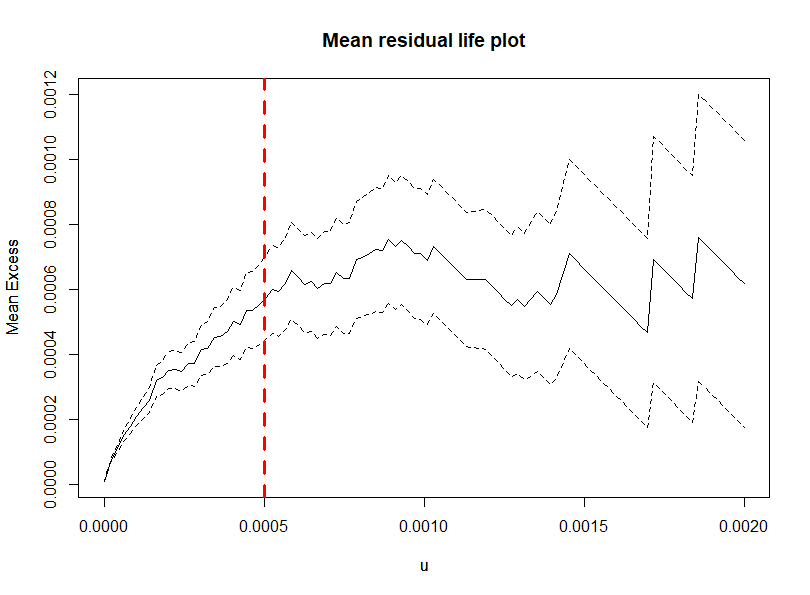}
\includegraphics[width= 0.48\textwidth]{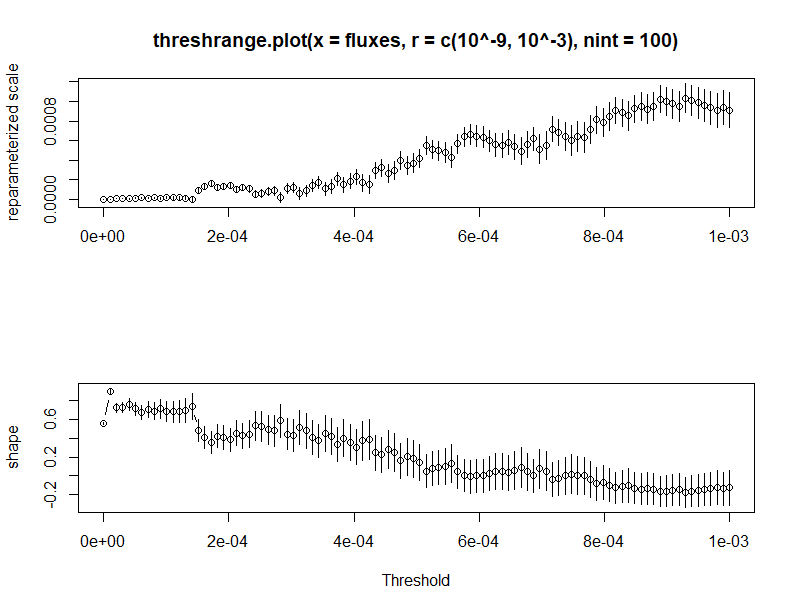}
\caption{Mean exceedance and threshold stability plot for the GOES data. The first plot shows the mean exceedances for a given threshold value versus the threshold $u$. The dotted lines represent the 95\% confidence intervals. One notes that linearity roughly starts after $u=5 \times 10^{-4}$ i.e. after an X5 flare. In the second plot, in the top row we show the scale $\sigma$ versus the threshold $u$ and in the bottom plot we have the shape $\xi$ versus $u$. Stability of both parameters with respect to the threshold starts at about $u=5 \times 10^{-4}$ for both plots thus making this choice valid.}
\label{fig:means}
\end{figure}

The first step is to estimate the threshold $u$ that determines the extreme values of our data. To determine that, as mentioned before, one needs to examine the mean residual life plot looking for linearity as well as the plot of the GPD parameters against a series of thresholds $u$ looking for stability. Figure \ref{fig:means} shows these plots where the mean residual life plot becomes linear and the threshold range plot becomes stable for a threshold value of $u = 5 \times 10^{-4}$ i.e. for X5 flares. For the rest of the statistical analysis, the value of $u=5 \times 10^{-4}$ is used as the threshold value as suggested by the diagnostics. According to NOAA \citep{NOAA} a flare of intensity X5 is classified between strong and severe so using this value as a threshold balances the ``extremity'' of events and having enough data points so that the estimated results are valid. The chosen threshold value gives a set of 93 exceedances i.e. 93 flares are greater than X5 in our given dataset. The Maximum Likelihood Estimates (MLE) and standard errors for the distribution parameters are: $\hat{\sigma} = 5 \times 10^{-4} \pm 0.2 \times 10^{-7}$ and $\hat{\xi} = 0.12 \pm 0.09$ which were found by using R and the package ``extRemes2'' \citep{extremes}. In general, the normal approximation to the distribution of the maximum likelihood estimator may be poor and more accurate confidence intervals are extraced from the so-called profile likelihod \citep{Coles}. With this approach, the estimate of $\xi$ suggests an unbounded distribution ($\xi >0$) and the evidence is reasonably strong as the 95\% profile likelihood confidence interval (C.I.) is almost exclusively in the positive domain with $\hat{\xi} \in (-0.017, 0.3589)$.

\begin{figure}[h!]
\hskip-1.5cm
\includegraphics[width= 0.6\textwidth]{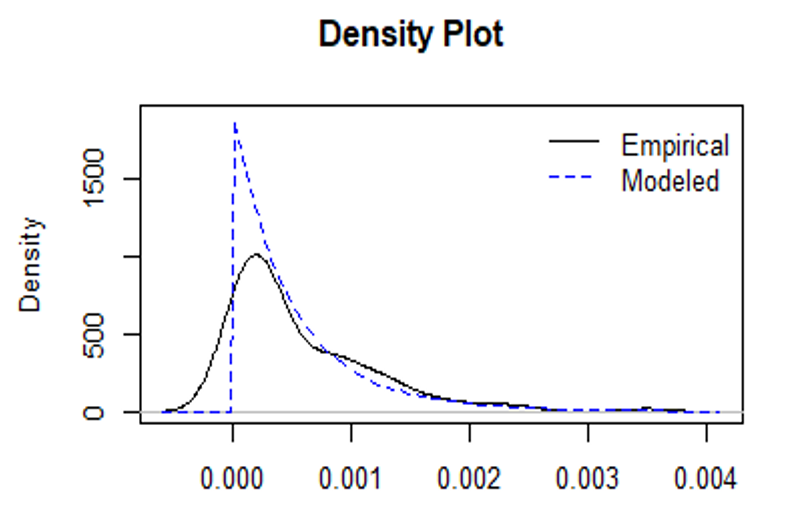}
\includegraphics[width= 0.48\textwidth]{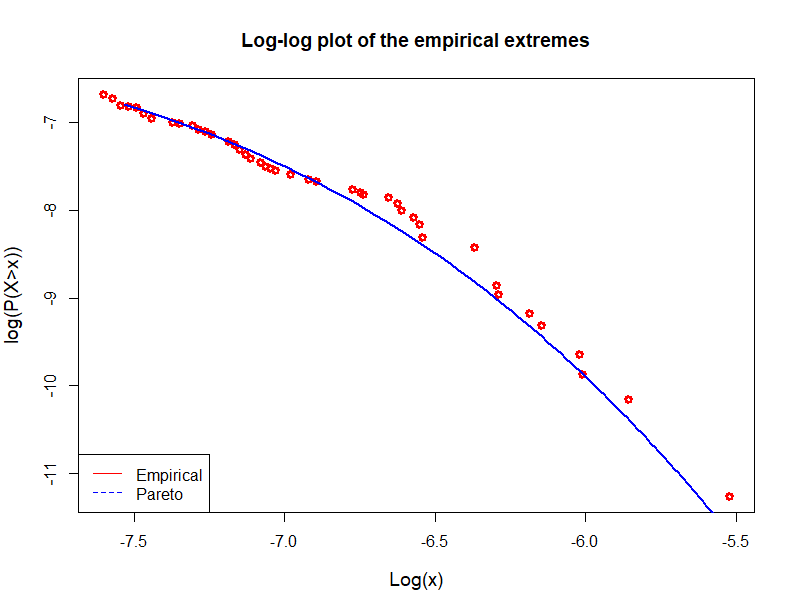}
\caption{Diagnostic plots. The first plot shows both the empirical and the fitted probability distribution whereas the second plot shows the empirical and predicted probabilities of extreme events on a log-log scale.}
\label{diagnostics}
\end{figure}

\begin{figure}[h!]
\centering
\includegraphics[width= \textwidth]{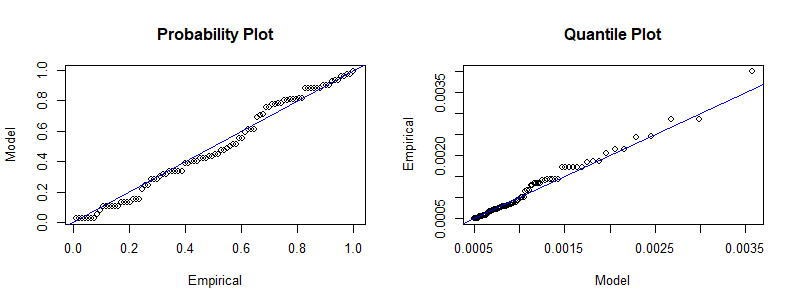}
\caption{Further diagnostic plots, PP and QQ-plots respectively. The goodness of fit validates the method used as in both plots the empirical model plotted against the theoretical falls onto an approximate straight line.}
\label{qq}
\end{figure}

\bigskip
\noindent
The validity of the model can be tested by the following diagnostic plots. From Figure \ref{diagnostics} we can confirm the goodness of fit of the GPD for all solar flares above the chosen threshold - both diagnostic plots appear reasonable fits to the observed data, especially the log-log plot indicating that our model predicts extreme events accurately. Furthermore, the PP-plot and the QQ-plot shown in Figure \ref{qq} look approximately linear, as expected when empirical predictions agree with the theoretical ones.

\begin{figure}[h!]
\centering
\includegraphics[width= 0.8\textwidth]{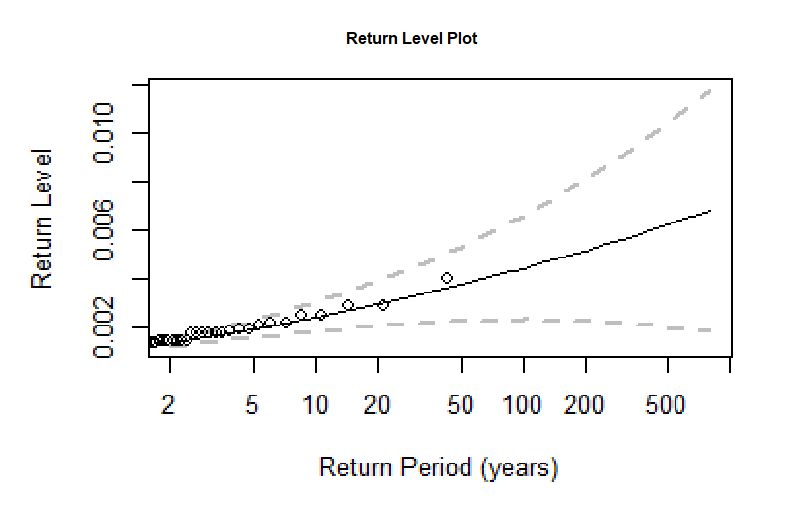}
\caption{Return levels and their 95\% confidence intervals (C.I.). A ``Halloween'' storm flare (X35) is expected once every 38 years whereas a Carrington event (X45) is expected once every 110 years.}
\label{return}
\end{figure}

\bigskip
\noindent
Since the model used is a good fit, it can be further used to estimate the extreme expected flares in a given period - this can be given by so-called \textbf{return levels}. The N-year return level is expected to be exceeded on average once every $N$ years. More precisely, $z_N$ is exceeded by the annual maximum in any particular year with probability $1/N$. This means that the $N$-year return level is given by:

\begin{linenomath*}
\begin{equation}
z_N = u + \frac{\sigma}{\xi} \left[ (N n_y \zeta_u)^{\xi}- 1 \right]
\end{equation}
\end{linenomath*}

\noindent
where $N$ is the number of years, $n_y$ the number of observations per year, $\zeta_u = \mathbb{P}(X>u)$ and $\xi$ the estimated shape parameter of the GPD. Using the return levels we can estimate Carrington-like events and their 95\% confidence intervals. These can be found in Table \ref{tab}, and a plot of these levels is provided in Figure \ref{return}. Caution is required in the interpretation of return level inferences, especially for return levels corresponding to long return periods since this corresponds to extreme extrapolation. Again better approximations are generally obtained from the appropriate profile likelihood function  which is used in the results of Table \ref{tab}. Though the EVT model is supported by mathematical arguments, its use in extrapolation is based on unverifiable assumptions, and measures of uncertainty on return levels should properly be regarded as lower bounds that could be much greater if uncertainty due to model correctness were taken into account.

\bigskip
\noindent
According to the estimated return levels, based upon the data analysed here, a Carrington-like event (X45) is expected as a one in 110-year event whereas a ``Halloween''-like event (X35) is a one in 38-years event. These have relatively large uncertainties as indicated by the confidence intervals (C.I). A Carrington-like event is expected in the next decade with probability 9\% whereas the probability of a Halloween-line event happening in the next decade is 23.8\%. Comparing these results to those carried out by \citep{Riley2017}, we can say that they are comparable although a Carrington-like event is predicted in the next decade with probability 10.3\%. This could be an indication of the overprediction caused by the use of power law models.

\begin{table}
\caption{Estimates and 95\% C.I. of several return levels}
\label{tab}
\centering
\begin{tabular}{|c|c|c|}  \hline 
 \large \textbf{Return level} & \large \textbf{Estimate} & \large  \textbf{ C.I.}   \\ \hline
  \large \textbf{11-year} &  \large  \textbf{X24.5} & \large (\textbf{X20.5}, \textbf{X35.5}) \\ \hline
  \large \textbf{20-year} & \large \textbf{X29.5} & \large  (\textbf{X23.5}, \textbf{X47.5}) \\ 
  \hline
\large   \textbf{38-year}  & \large  \textbf{Halloween} (\textbf{X35})  & \large (\textbf{X26.5}, \textbf{X65}) \\  \hline
\large \textbf{50-year}  & \large  \textbf{X37.5}  & \large (\textbf{X27.5}, \textbf{X74}) \\  \hline
\large \textbf{100-year}  &  \large \textbf{X44} & \large  (\textbf{X30.5}, \textbf{X103}) \\\hline
\large   \textbf{110-year}  & \large \textbf{Carrington} (\textbf{X45}) & \large (\textbf{X31}, \textbf{X108}) \\  \hline
\large \textbf{150-year}  &  \large \textbf{X48} & \large  (\textbf{X32.5}, \textbf{X125}) \\\hline
\end{tabular}
\end{table}

\begin{figure}[h!]
\hskip-0.8cm
\includegraphics[width= 0.5\textwidth]{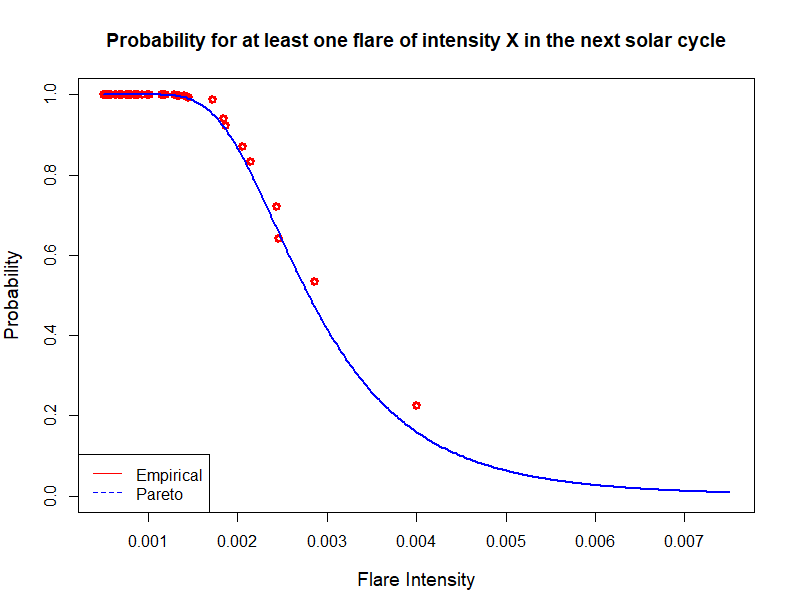}
\includegraphics[width= 0.5\textwidth]{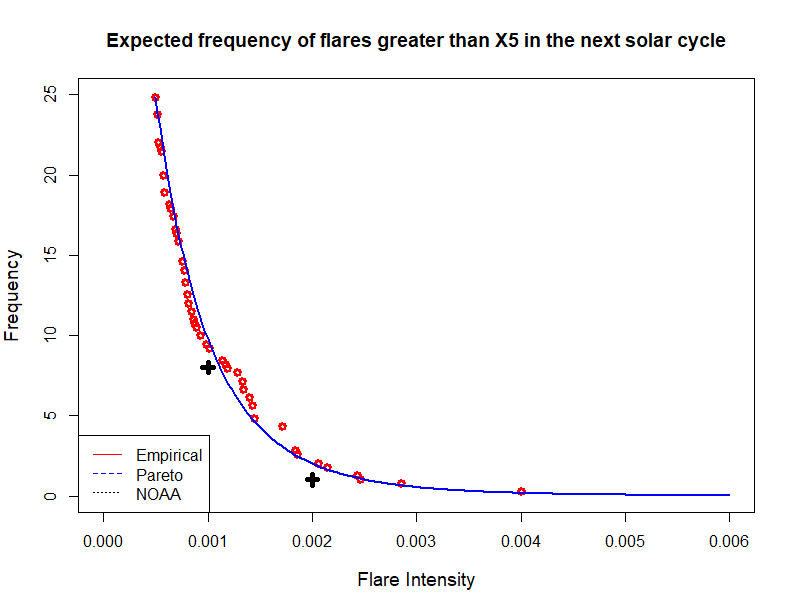}
\caption{The first plot shows the empirical expected frequency of a flare of certain intensity in the next solar cycle. The expected frequency is predicted well by the model --  especially the tail -- whilst our predictions are in close proximity to the predictions provided by NOAA. The second plot shows the probability of having a flare of certain intensity in the next solar cycle.}
\label{freq}
\end{figure}

\bigskip
\noindent
In Figure \ref{freq}, the first plot demonstrates the empirical and the fitted probability of at least one flare of a certain intensity happening in the next solar cycle. One can see that this is a relatively good fit to the data and it has been estimated as follows. Let $p = \mathbb{P}(X>x)$ be the unconditional probability of a single flare being greater than $x$. The average number of flares  in the solar cycle is $n_y \times 11$, with $n_y=1799$ the average number of observations per year. If $\mu$ is the number of flares in a cycle whose intensities are greater than $x$ then $\mu$ is a random variable with a binomial distribution with parameters $n_y \times 11$ and $p$. The probability that in a cycle we observe at least one flare greater than $x$ is:

\begin{linenomath*}
\begin{equation}
\mathbb{P}(\mu>x) = 1- (1-p)^{n_y \times 11}
\end{equation}
\end{linenomath*}

\noindent
To lend further weight to the argument, the second plot of Figure \ref{freq} demonstrates the empirical and the fitted expected frequencies of a flare of a certain intensity in a solar cycle. These were estimated as the expected value from the binomial distribution described above and thus is equal to $p \times n_y 11$. One can see from this plot that the tail is predicted very well by the model. Furthermore, it is shown that our results are in close agreement with NOAA's frequency predictions.

\section{Seasonality}
\label{seasonality}

It has been suggested that seasonality inherited from the solar cycle affects the frequency as well as the severity of the solar flares emitted. To investigate this the solar seasonality has been subtracted from the data in the following way: a periodogram was used to identify the dominant frequencies of the discrete Fourier transform of the data, $\omega_{1}$ and $\omega_{2}$, which were found to correspond to periods of 11 and 14 years respectively. Sinusoidal waves of the form $\sum_{i=1}^{n=2} A_{i} \sin(\omega_{i} t)+ B_{i} \cos(\omega_{i} t) + C$, were fitted to the data using a nonlinear least-squares estimation procedure (``nls'' in R). The estimated function was subtracted from the data and the analysis we used in Section 5 was repeated. We found the effects of seasonality to be negligible with regard to extreme events: the estimates of the distribution parameters of the de-seasonalised data were almost equivalent to those reported earlier and were perfectly compatible taking their confidence interval into account. This suggests that solar seasonality does not significantly impact on extreme events and thus their prediction.

\section{Conclusions}
\label{conclusions}

EVT is a rigorous method to estimate the intensity of solar flares happening years into the future and acknowledges the errors in extrapolation in the estimates in the big limit. We have shown that the use of EVT is consistent with the data provided by SWPC.

\bigskip
\noindent
The fit to the extremes of the sample distribution function is excellent and the results show that the extreme events observed are well fit by the EVT models. EVT also provides us with results that can be interpreted as the worst case scenario i.e. as the most extreme events in a given period of time. For example the return period of a Carrington-like event (X45) is 110-years and a ``Halloween''-like event (X35) is 38-years. Moreover, for the purposes of anticipatory planning to cope with potential extreme events, it would be recommendable to take the prediction of the mode pessimistic value within confidence intervals for the most secure preparation. Moreover, the probability of a Carrington-like event happening in the next decade is 9\% whereas a Halloween-like event is expected in the next decade with probability 23.8\%. These two results are compatible with previous predictions \citep{Riley2017} and could be an indication of the overprediction caused by the use of power law models. 

\bigskip
\noindent
Furthermore, the model used seems perfectly valid for the frequency distribution of the empirical data; the predictions conform to those provided by NOAA \citep{NOAA}. EVT is a great tool for the estimation of values that have not been observed and as more data are collected, the estimates of the parameters approach more closely the true parameters of the extreme value distribution function. Since the model seems appropriate for the whole tail of the data it could suggest that one mechanism of energy release is sufficient to produce solar flares across the whole range of strengths which might not be the case were we to have seen significant deviation from the tail of our fitter model for large observed flares. 




\bigskip
\noindent
The results show that taking into account the data from the last four solar cycles an X24 flare is expected once every 11 years. However in this solar cycle (24) the most energetic flare recorded was X9.3 on September 6, 2017. The amplitude of the solar cycle 24 was lower than previous cycles and as we observed it produced fewer energetic events. The influence of the amplitude of each solar cycle in the confidence to forecast in the long term requires a better understanding of the particularities of each sub set of data (by cycle) and their relation with global statistics.  

\subsection*{Acknowledgements}
\noindent
Both authors acknowledge access to solar flare's flux strength data provided by NOAA's National Centers for
Environmental Information website found at \url{https://www.ngdc.noaa.gov/stp/space-weather/solar-data/solar-features/solar-flares/x-rays/goes/xrs/}. TT would like to thank J.P. Edwards for fruitful discussions whilst writing this paper. 


\end{article}

\end{document}